\documentclass[aps,english,prb,twocolumn,showpacs,superscriptaddress]{revtex4-1}
\usepackage{graphicx}

\begin{document}

\title
{Superlubricity through graphene multilayers between Ni(111) surfaces}

\author{S. Cahangirov}
\affiliation{UNAM-National Nanotechnology Research Center, Bilkent University, 06800 Ankara, Turkey}
\affiliation{Institute of Materials Science and Nanotechnology, Bilkent University, Ankara 06800, Turkey}

\author{S. Ciraci}\email{ciraci@fen.bilkent.edu.tr}
\affiliation{UNAM-National Nanotechnology Research Center, Bilkent University, 06800 Ankara, Turkey}
\affiliation{Institute of Materials Science and Nanotechnology, Bilkent University, Ankara 06800, Turkey}
\affiliation{Department of Physics, Bilkent University, Ankara 06800, Turkey}

\author{V. Ongun \"{O}z\c{c}elik}
\affiliation{UNAM-National Nanotechnology Research Center, Bilkent University, 06800 Ankara, Turkey}
\affiliation{Institute of Materials Science and Nanotechnology, Bilkent University, Ankara 06800, Turkey}

\begin{abstract}

A single graphene layer placed between two parallel Ni(111) surfaces screens the strong attractive force and results in a significant reduction of adhesion and sliding friction. When two graphene layers are inserted, each graphene is attached to one of the metal surfaces with a significant binding and reduces the adhesion further. In the sliding motion of these surfaces the transition from stick-slip to continuous sliding is attained, whereby non-equilibrium phonon generation through sudden processes is suppressed. The adhesion and corrugation strength continues to decrease upon insertion of the third graphene layer and eventually saturates at a constant value with increasing number of graphene layers. In the absence of Ni surfaces, the corrugation strength of multilayered graphene is relatively higher and practically independent of the number of layers. Present first-principles calculations reveal the superlubricant feature of graphene layers placed between pseudomorphic Ni(111) surfaces, which is achieved through the coupling of Ni-$3d$ and graphene-$\pi$ orbitals. The effect of graphene layers inserted between a pair of parallel Cu(111) and Al(111) surfaces are also discussed. The treatment of sliding friction under the constant loading force, by taking into account the deformations corresponding to any relative positions of sliding slabs, is the unique feature of our study.

\end{abstract}

\maketitle

\section{Introduction}

When placed between two strongly interacting surfaces, inert substances such as atoms or molecules reduce the adhesion and sliding friction by screening the interaction between them. Progress in atomic scale sliding friction\cite{persson,mate,tomanek,buldum1,gnecco,meyer,review,buldum2,renaissance,seymur} and molecular lubrication\cite{persson,buldum} have made lubricant materials an intense field of research in nanotribology. Layered materials composed of weakly interacting two-dimensional (2D) single layers, such as molybdenum disulfide (MoS$_2$) and graphite were used as solid lubricants in diverse applications long before single layers of these materials were isolated. The key feature which make these materials so important in tribology is their strong covalent intralayer bonds in contrast to the weak van der Walls (vdW) interlayer interactions. Recently, the contrast between these intralayer and interlayer interactions were quantified in terms of frictional figure of merit and it was predicted that WO$_2$ can show better lubricant performance as compared to MoS$_2$.\cite{seymur,mx2}

Graphene, the single layer of graphite, with its exceptional physical and chemical properties, has also been a subject of interest for tribological applications.\cite{dienwiebel,filippov,fasolino,lebedeva,anisotropy,popov} It has been shown that graphene layers can stick to metal surfaces and provide excellent protection from oxidation.\cite{topsakal} The friction can be further reduced by rotating sliding graphene layers relative to each other, so that they become incommensurate with a flat potential corrugation.\cite{dienwiebel,anisotropy}

Recent works have concentrated on how the friction force between the tip of a force microscope and graphene surface varies as the number of layers increases from single layer to multi layers representing graphite.\cite{lee, carpick, bennewitz} Lee \textit{et al.}\cite{lee} showed that the friction force between SiN tip and graphene flake prepared on silicon oxide monotonically decreases as the number of graphene layers are increased. A similar trend observed also for different tip material and substrate holding graphene was attributed to the reduced piling or puckering of the layers with increasing number of layers.\cite{carpick} In contrast, Filleter \textit{et al.}\cite{bennewitz} found that friction force is higher on graphite compared to bilayer graphene, since electron-phonon coupling is suppressed in the latter. Kim \textit{et al.}\cite{kim} reported that the adhesion and friction coefficient between SiO$_2$ lens and graphene deposited SiO$_2$ substrate are reduced.

In this paper we investigate the interaction, the strength of the potential corrugation and energy dissipation between two Ni(111) surfaces having $n$ layers ($n=0-5$) of graphene in between. Our main objective is to reveal the physics of interactions pertaining to the lubrication capacity of graphene as a prototype for similar single-layer nanomaterials. In this respect our focus is different from previous experimental studies which are dealing with the sliding friction between a tip and graphene layers. The present approach mimics a realistic situation where the metallic surfaces are coated by lubricant layers and the radii of asperities are much larger when compared with atomic scales. In order to hinder other effects such as size, edge, rippling and incommensurability from interfering in our analysis, we treat large surfaces in terms of periodically repeating primitive unit cells using periodic boundary conditions. This way our model is isolated from these stochastic effects to reveal the physics underlying more fundamental and material specific interactions. The nature of interaction between sliding bare Ni(111) surfaces and those between graphene layers which cover Ni(111) surfaces, as well as lateral forces generated therefrom necessitate a quantum mechanical treatment of the sliding phenomenon. Thus we carried out calculations using quantum mechanical methods as 2D layers execute a 3D sliding motion under a given constant normal force. The sliding motion under the constant force mode, where the structure is optimized for any relative positions of slabs is the crucial and unique aspect of the present study based on first-principles Density Functional Theory (DFT).\cite{dft-1,dft-2}

\begin{figure}
\includegraphics[width=8.5cm]{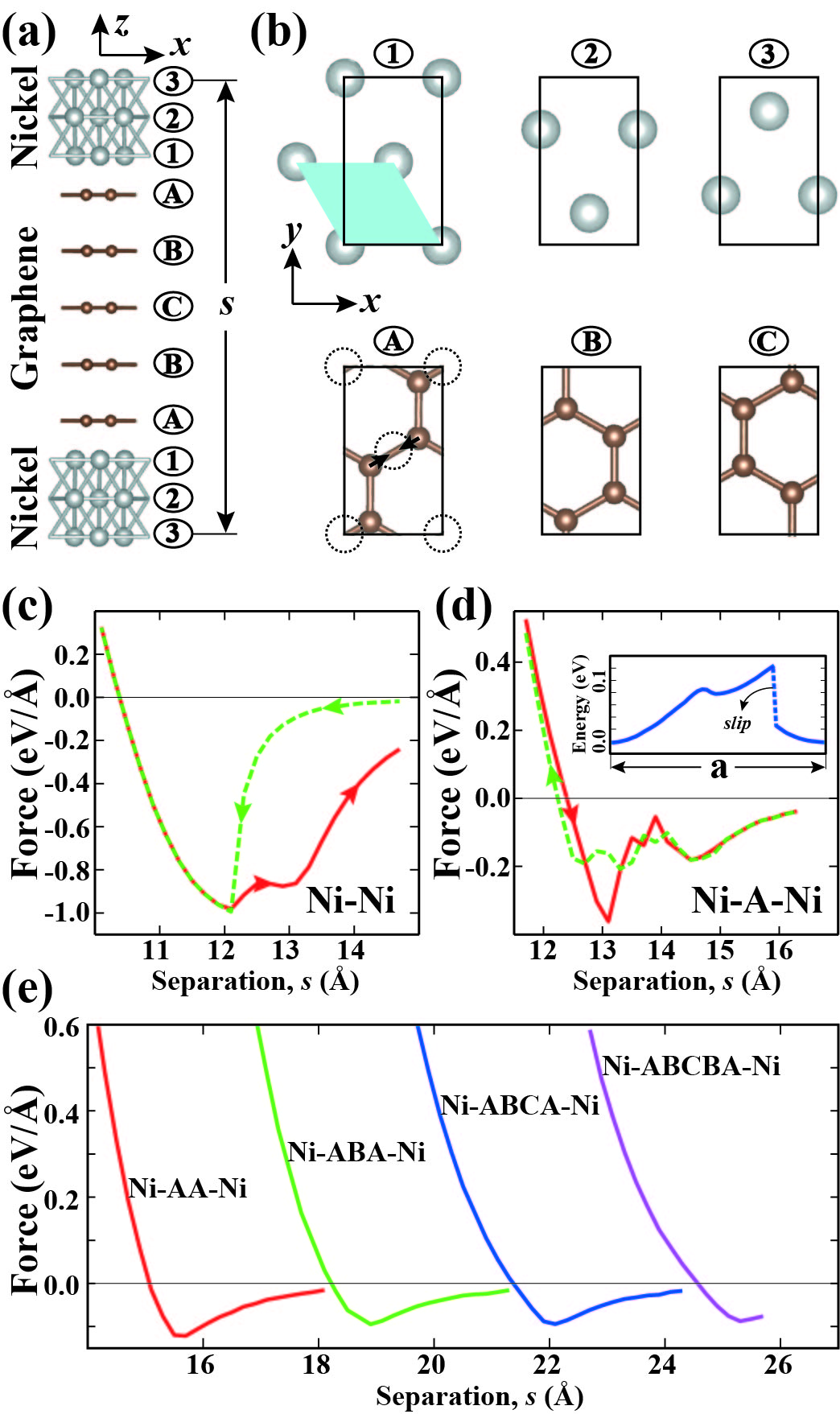}
\caption{(a) Side view of the arrangement of the Ni-ABCBA-Ni structure. The outermost Ni(111) atomic planes are fixed at the separation $s$. (b) Top view of individual layers constructing the Ni-ABCBA-Ni structure. The primitive unitcell of graphene is shown by blue shaded area. Dotted circles represent optimized positions of Ni atoms below the graphene layers in configuration $A$. Adhesion hysteresis curves for Ni-Ni in (c), and for Ni-A-Ni structures in (d) and its stick-slip behavior shown by inset. (e) Normal force along $z$ axis $F_{z}$ as a function of separation $s$ for Ni-graphene-Ni structures with $n$=2-5 graphene layers. The unit of $F_z$ is eV/\AA~per unit cell. Ni and carbon atoms are indicated by large/blue and small/braun balls, respectively.}

\label{fig1}
\end{figure}

We highlight our study, which starts with an extensive analysis of the Ni(111) slab coated with single layer graphene as follows: (i) Strong adhesive forces which lead to strong energy dissipation and wear are dramatically reduced even when only one single layer of graphene is inserted between two Ni(111) surfaces. (ii) Even more remarkable is that after the second layer of graphene is inserted, each layer is attached to one of metal surfaces and adhesion is further reduced. At the end, the stick-slip regime and hence phononic energy dissipation is suppressed and the system enters into the continuous sliding regime. (iii) By inserting more graphene layers between Ni(111) surfaces the corrugation strength decreases gradually and saturates at a fixed value. (iv) On the other hand, if the supporting metal surfaces are removed, the friction between graphene layers sliding on top of each other are relatively larger and practically independent of the number of layers $n$ in between. The above features leading to nearly frictionless sliding are specific for Ni(111) surface which is almost lattice matched (or pseudomorphic) to graphene, and originate from a special coupling between Ni-$3d$ and graphene-$\pi$ orbitals accompanied with a complex charge exchange between them. This special interaction also explains why the growth of graphene on Ni(111) surface with rather low barriers for defect healing is favored.\cite{lahiri,ongun} The reduction of adhesion and sliding friction by graphene layers placed between the pairs of bare insulator surfaces and the (111) surfaces of metals, such as Al, Cu and Ni studied here show that the interaction between graphene and Ni(111) surface appears to be rather unusual. In the rest of the paper, superlubricity due to graphene will be investigated for Ni(111), while the situations with other metal surfaces, such as Cu(111) and Al(111) will be discussed.

\section{Method}
We performed first-principles plane wave calculations within the Generalized Gradient Approximation (GGA) \cite{pbe} including van der Waals corrections (vdW)\cite{vdw} using PAW potentials \cite{paw}. A plane-wave basis set with kinetic energy cutoff of 500 eV is used. In the self-consistent potential and total energy calculations the Brillouin Zone is sampled by (13$\times$13$\times$1) k-points. For every perpendicular and lateral configuration of slabs the equilibrium positions of the metal and graphene atoms are obtained by optimizing all atomic positions and lattice constants. This way static deformations under perpendicular loading force are taken into account. The total energy and atomic forces are minimized by using conjugate gradient method. The convergence for energy is chosen as 10$^{-5}$ eV between two steps, and the maximum force allowed on each atom is less than 10$^{-4}$ eV/\AA. Numerical plane wave calculations have been performed by using VASP package.\cite{vasp1,vasp2} Despite its limitations in the excited state properties, DFT calculations have provided crucial contributions to our understanding of graphene based materials. In particular, atomic structure and mechanical properties have been revealed with reasonable accuracy, when suitable approximation is made for exchange-correlation potential and caution is exercised in structure optimization. The capacity of DFT will be elaborated in section IIIA.

\subsection{Atomistic Model}
The adhesion and frictional properties of graphene layers sandwiched between two metal surfaces and those of bare graphenes are treated using the models specifically described for Ni(111) in Fig.~\ref{fig1}. Two metal parts executing relative motion in the perpendicular or lateral directions are represented by two slabs each consisting of three (111) atomic planes of metals. We apply periodic boundary conditions with the unit cell comprising one metal atom in each (111) metal plane and two carbon atoms for each graphene layer. Along the $z$-direction, which is perpendicular to the surfaces (or to the $xy$-plane), the interaction between periodic images of Ni slabs is hindered by introducing a vacuum spacing of 15~\AA. The structure presented in Fig.~\ref{fig1}(a) is named as Ni-ABCBA-Ni, where A, B, C,.. are graphene layers corresponding to equilibrium in-plane configuration of carbon atoms. To avoid any confusion, the atomic layers comprising the Ni slabs are arranged in a mirror symmetry. This arrangement is presented in Fig.~\ref{fig1}(a), while the configuration of Ni and carbon atoms in each plane is shown in Fig.~\ref{fig1}(b). In the optimized structure, Ni atoms, which are positioned at the bridge sites of graphene structure attract carbon atoms and slightly break the hexagonal symmetry, as shown in Fig.~\ref{fig1}(b).

\section{Computational Results}
Superlubricity through graphene layers between sliding Ni(111) surfaces is investigated in the following sequence: We first examine the energetics and atomic configuration of the composite material consisting of single graphene adlayer attached to the Ni(111) slab. This configuration represents the coating of Ni(111) surface. Secondly, we constructed the optimized structures of Ni-A...-Ni model and investigated the character of interaction energy and force between two Ni slabs with and without graphene layers between them. We calculated the variation of adhesion forces while approaching and pulling two slabs in Fig.~\ref{fig1}(a). In general, adhesion can be taken as the measure of the sliding friction between two surfaces in relative lateral motion. Finally, nearly frictionless sliding and the superlubricity through graphene layers between metal surfaces are investigated in the constant force mode.

\subsection{Single Graphene adlayer on Ni(111) Slab}
Ni(111) surface is special mainly for two reasons: (i) Ni(111) surface is almost lattice matched to graphene. We predict the optimized lattice constants of free graphene and free Ni(111) slab consisting three Ni(111) atomic planes as $a_{Gr}$=2.47 \AA~ and $a_{Ni}$=2.43 \AA, respectively. The distance between Ni(111) layers are found to be $d_{Ni}$=2.02 \AA. The percentage difference between their optimized lattice constants, i.e. $\Delta a =(a_{Gr}-a_{Ni})/a_{Gr}$ is only 1.6\%. Investigating whether this small difference can be compensated between graphene and Ni(111) surfaces, or misfit dislocations are generated to relieve the strain requires very accurate DFT calculations comprising 20.000 C and Ni atoms, which cannot be achieved in terms of the present first-principles approach. Under these circumstances, we calculated the total energy of graphene adlayer on Ni(111) slab consisting of three Ni atomic planes by optimizing their lattice parameters in the common primitive unit cell. This reflects the actual situation, where only a few Ni layers at the surface can be deformed due to graphene adlayer. However, the deformation ceases and Ni recovers its equilibrium structure, when one goes away from the interface.  Upon optimization of composite structure graphene+Ni(111) slab, the lattice constant of free graphene is compressed to 2.45 \AA, while Ni(111) slab is extended to compensate the strain energy. Because of the expansion of Ni slab, its interlayer spacing decreased from 2.02 \AA~ to 2.00 \AA. The optimized interlayer spacing between graphene and Ni(111) slab is crucial for our study and is calculated to be 2.05~\AA. As described in Fig.~\ref{fig1}(b), the positions of carbon atoms relative to Ni atoms below in the first layer of Ni slab corresponds to the top-bridge site.\cite{cabrera} The binding energy $E_b$ between graphene adlayer and Ni(111) slab is obtained from the following expression, $E_b = E_{T}[Gr+Ni]-E_{T}[Gr]-E_{T}[Ni]$ in terms of the optimized total energies of Graphene+Ni combined structure $E_{T}[Gr+Ni]$, single layer, free graphene $E_{T}[Gr]$ and free Ni slab $E_{T}[Ni]$, respectively. We found the binding energy to be 264 meV per primitive unit cell or 51 meV/\AA$^{2}$.

In regard to the single graphene adlayer on Ni(111) following comments are in order: (i) In layered structures like graphite and h-MoS$_2$, the vdW interaction has a significant contribution to the binding of layers. The interaction between layers have been examined using different exchange-correlation approximation with and without vdW correction.\cite{topsakal5} It has been revealed that LDA and GGA(PW91)+vdW provides good predictions for the lattice parameters, in particular for the interlayer separations of graphene and MoS$_2$. Apparently, LDA overbinds in spite of the fact that it does not include vdW interaction. Earlier, the structural parameters and binding energy of graphene adlayer on Ni(111) surface have been investigated also by different approximations leading to different values.\cite{cabrera,buehler,mortensen} We believe that the GGA(PBE) with vdW correction\cite{vdw} used in our calculations appears to be the most physical approximation. The present prediction for the spacing between graphene-Ni(111) surface is   2.05~\AA~ is in good agreement with the experimental value\cite{expd} of 2.11~\AA. Our result regarding the spacing between graphene adlayer and Ni(111) differs from the previous study\cite{mortensen} due to perhaps different schemes used for vdW correction. The vdW correction used in our study\cite{vdw} have been successful in various graphene and metal based systems. (ii) In our study, the optimization of lattice parameters using the conjugate gradient has started with different values of the lattice spacing to prevent the structure from trapping at spurious minima. Earlier, the theoretical studies used lattice constants of Ni(111) under graphene that obtained either from PBE calculations\cite{cabrera}, or from experiment\cite{buehler}. The present optimized bridge-top configuration is in agreement with earlier LDA result.\cite{cabrera} That the graphene adlayer on Ni(111) surface is predicted to be non-bonding in a previous study\cite{cabrera} indicates the importance of the vdW interaction and hence corroborates our approach.  (iii) We believe that DFT calculations have been useful in understanding various graphene based systems as long as calculations are carried out using appropriate approximations by optimization of the atomic structure and lattice parameters with care and taking into account the weak van der Waals interactions properly. It should be noted that the calculated atomic and electronic structures may vary depending on the approximation used for the exchange correlations.

\begin{figure}
\includegraphics[width=8.5cm]{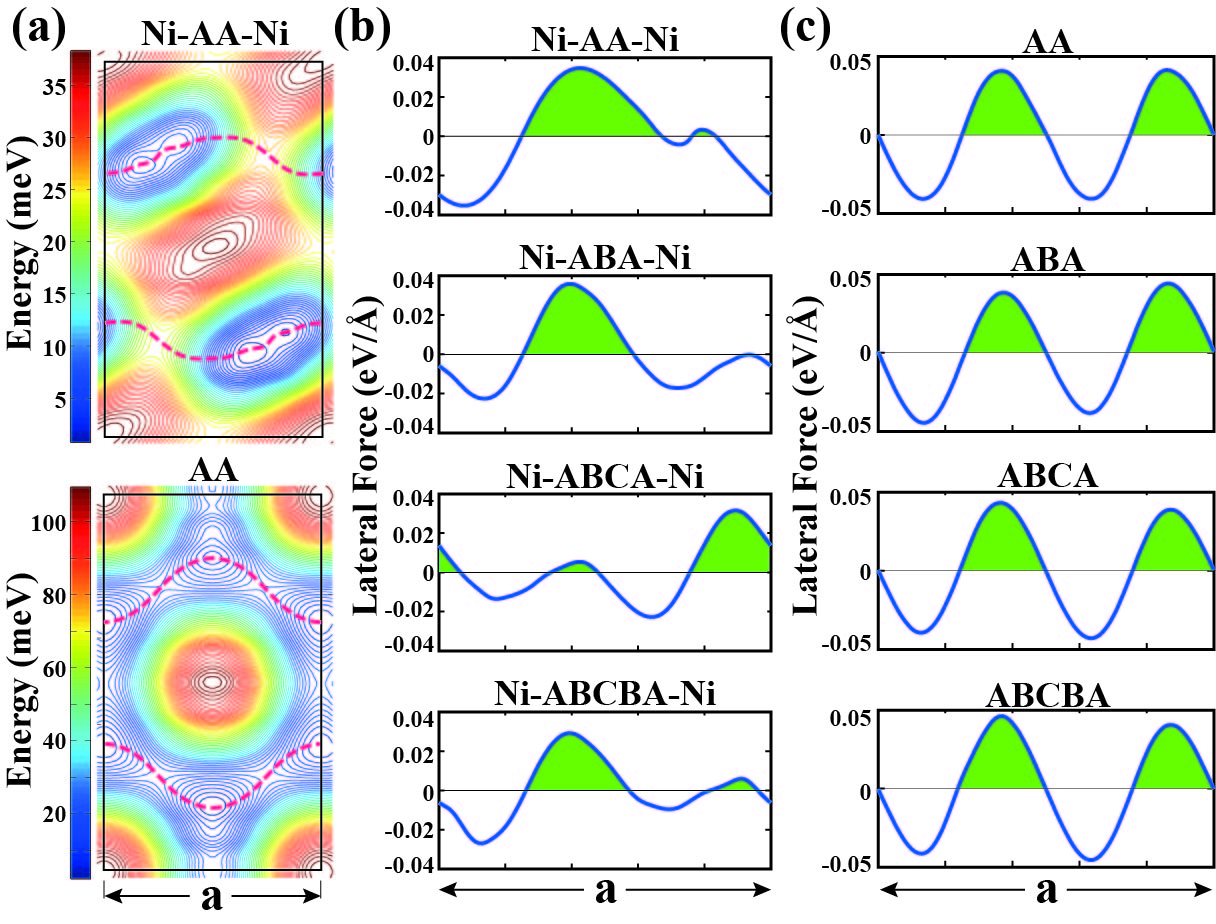}
\caption{(a) Profiles (contour plots) of potential corrugation for Ni-AA-Ni and AA [without Ni(111) slabs] structures calculated under constant pressure of 7~GPa. The paths along which one slab moves in the course of sliding when pulled along $x$ axis are shown by red dashed lines. The lattice constant of the unit cell is indicated by \textbf{a}. (b) Variation of lateral force $F_{x}$ along $x$-axis during sliding of Ni-AA-Ni structures over the path shown in (a). The integral of shaded (green) areas is defined as the corrugation strength $W_D$ (see text). (c) Same as (b) for sliding AA structures without Ni(111) slabs.}

\label{fig2}
\end{figure}

\subsection{Adhesion between surfaces}
The analysis adhesion starts by calculating the normal force $F_z$ at outermost atoms of Ni slabs, which are kept fixed in the course of relaxation for a given separation, $s$ as described in Fig.~\ref{fig1}(a). The force variation is obtained while gradually varying $s$ in small steps. We start by two Ni slabs with no graphene layer in between. The dashed curve in Fig.~\ref{fig1}(c) shows the variation of $F_z$ as $s$ is gradually decreased. One can observe a slightly attractive region followed by a sudden increase in the attractive force. This is the phenomenon known as the jump-to-contact. During the sudden increase of normal force in the attractive range, both Ni slabs are expanded towards each other and after this stage the distance between the facing atomic layers remains nearly constant until equilibrium is reached. When $s$ is further decreased, the attractive force decreases and eventually becomes repulsive.

In Fig.~\ref{fig1}(c) the normal force $F_{z}$ follows a different route as $s$ increases (or as one of Ni(111) slabs is pulled-off) starting right at the minimum point of $F_{z}(s)$ curve. The observed bistability or hysteresis manifests itself as strong adhesion and wear phenomena frequently observed in the dry sliding friction between two metal surfaces. Similarly, the normal forces between two bare Cu(111) and Al(111) slabs are strong and show the behavior similar to Fig.~\ref{fig1}(c).

Next, we consider one single graphene layer, which is inserted between two Ni slabs at the minimum energy configuration $A$ described in Fig.~\ref{fig1}(b). The maximum attractive force between Ni(111) surfaces in Fig.~\ref{fig1}(d) is reduced to approximately 1/4 of that between bare surfaces. Even if this graphene layer screens (or decouples) the interaction between two bare Ni(111) surfaces and hence dramatically decreases the maximum attractive force between them, the hysteresis is still present between the approach and pull-off of the top surface, as seen in  Fig.~\ref{fig1}(d). The sudden variation of energy in the constant height sliding mode is also shown by inset. This clearly demonstrates that the stick-slip motion, which usually plays the principal role in the dissipation of mechanical energy is still present. A pair of Cu(111) and a pair of Al(111) slabs having single graphene layers in between also exhibit similar behaviors, such as the reduced adhesion and bistability occurring in the course of the pushing and pulling-off the metal slabs. This is the first important theoretical prediction of our analysis and explains why graphite flakes can provide the lubrication of the sliding motion of two parallel metal surfaces.

A number of interesting effects occur when the second graphene layer is inserted between Ni slabs: Owing to significant graphene-Ni attraction each layer becomes stuck to one of the Ni(111) surfaces. Later we show that the binding between Ni(111) surface and graphene layer is achieved mainly through the coupling between Ni-$3d$ and graphene-$\pi$ orbitals. Under these circumstances the hysteresis is completely removed, since the jump-to-contact between graphene coated Ni slabs are hindered. Also the maximum attractive force between surfaces in Fig.~\ref{fig1}(d) is further reduced to $\sim$0.12 eV/\AA~per unit cell as shown in Fig.~\ref{fig1}(e). Consequently, the stick-slip regime comes to an end with the onset of the the continuous sliding regime, whereby the energy dissipation through the generation of non-equilibrium phonons is, in principle, completely suppressed. Reminiscent of Ni-AA-Ni in Fig.~\ref{fig1}(e), our results obtained from Cu-AA-Cu and Al-AA-Al indicate further weakening of the maximum attractive force upon the insertion of second graphene, where one graphene layer sticks to each metal surface. This is due to the fact that attractive metal-graphene interaction is stronger than the graphene-graphene interaction. Physical mechanisms underlying these effects will be clarified in the forthcoming analysis.

Three graphene layers in Ni-ABA-Ni shows slight decrease of maximum attractive force by $\sim$0.04 eV/\AA~per unitcell. As seen in Fig.~\ref{fig1}(e), the maximum attractive force or the adhesion decreases gradually for $n >$2 and eventually saturates at a constant value of force $\sim$0.08 eV/\AA~per unitcell. This is in compliance with the short range nature of the orbital overlaps of different layers. In contrast, depending on the range and type of metal-graphene interaction, the maximum attractive force in Cu-ABA-Cu and Al-ABA-Al slightly increases as compared to Cu-AA-Cu and Al-AA-Al and remain practically unaltered for more graphene layers. We calculated the increase of the maximum attractive force upon increasing the number of graphene layers from $n$=2 to $n$=3 to be 0.13eV/\AA~and 0.06 eV/\AA~per unit cell for Cu and Al, respectively. The value of the maximum attractive force for $n >3$ keeps the same value. This is the marked difference between Ni(111) and other metal surfaces like Cu(111) and Al(111) surfaces.

\subsection{Superlubricity through Graphene Layers}  

To clarify how graphene layers function as lubricant and also to investigate the effect of including more layers on the friction we examine the potential corrugation in the course of the sliding of the layers under constant pressure. To this end, we first calculate the optimized total energies $E_T$, when the topmost Ni layer is displaced and kept fixed at various lateral $(x,y)$-position and vertical separation $s$ relative to the bottommost layer (see Fig.~\ref{fig1}(a)).\cite{seymur} These calculations are performed in a 3D grid of $x,y,s$. The intervals between the data points were taken to be $\sim$ 0.2~\AA~ in the lateral $(x,y)$-plane and 0.2~\AA~ in the perpendicular direction, $s$, which are then made finer down to $\sim$ 0.05~\AA~ by spline interpolation. We also generate $F_x$, $F_y$ and $F_z$ matrices from the gradient of the total energy $F_{x,y,z}=-\partial{E_T(x,y,z)}/\partial{x,y,z}$, which are found to be consistent with the forces calculated from the Hellmann-Feynman forces on fixed atoms of outermost planes. We then retrieve $F_x$ and $F_y$ corresponding to a given $F_z$ (normal pressure) at each $(x,y)$ in the unit cell and generate the profiles of the potential corrugation from $\int^{x,y,F_{z}} (F_{x}dx+F_{y}dy)$, where the minimum of total energy is set to zero. The profiles (contour plots) of potential corrugation calculated for  Ni-AA-Ni and AA i.e. two flat graphene layers without Ni(111) are shown as the top and bottom panels of Fig.~\ref{fig2}(a), respectively. We note that the amplitude of the potential corrugation (i.e.the difference between the minimum and maximum of energy) is an order of magnitude smaller compared to those between two sliding, single-layer honeycomb structures of graphane CH, fluorographene CF, MoS$_2$ and WO$_2$ discussed in Ref. [\onlinecite{seymur}]. On the other hand, the intrinsic stiffness of the present case, which is related to the interaction between Ni and graphene layers is also substantially lower ($k_{s}=0.8$~eV/\AA$^2$) compared to the intrinsic stiffness of those honeycomb structures.\cite{seymur} The lower intrinsic stiffness accompanied by low potential corrugation curvature results in a frictional figure of merit of $\sim$~10, at constant pressure of 7~GPa, which is enough to keep the system in the continuous sliding regime. Comparing the profiles of the potential corrugation of Ni-AA-Ni and AA structures, one can see how the interaction between graphene layers is affected by their interaction with Ni surfaces. The effect of distortion presented in Fig.~\ref{fig1}(b) is reflected to the potential corrugation of Ni-AA-Ni, since its symmetry is changed from hexagonal to rectangular.

To set a measure for the \textit{corrugation strength} we first derive the path on which the upper slab would slide if it was pulled along $x$-axis. This path is shown by dashed lines in Fig.~\ref{fig2}(a) for the case of Ni-AA-Ni. In the case of structures having more than two graphene layers the path is found directly by starting from the Ni slab positions presented in Fig.~\ref{fig1} and moving along the $x$-axis while minimizing the total energy along $y$-axis. Then we calculate the lateral force $F_x$ along $x$-axis felt by the slab, as shown in Fig.~\ref{fig2}(b). Here we note that in the sliding of Ni(111) slabs having $n$ graphene layers the dissipation of energy through non-equilibrium phonons generated by sudden processes is hindered for $n \geq 2$ and hence $W=\int_0^a F_x dx$ vanishes. This, however, does not precludes energy dissipation through other mechanisms. With a premise that the maximum of the energy to be dissipated by any mechanism should not exceed $W_D=\int_0^a F_x^> dx$ (i.e the integral of all positive work done during sliding of one slab over one unit cell shown by the green shaded region in Fig.~\ref{fig2}(b)), we took $W_D$ as a measure for the corrugation strength. Clearly, this a stringent criterion for the sliding friction. The result of these calculations are presented in Fig.~\ref{fig2}(b). Note that $W_D$, which is also related to kinetic friction coefficient $\mu_{k} = (W_{D}/a)/F_{z}$, is already very small. To check the effect of the type of stacking we have also calculated the force variation for Ni-ABABA-Ni structure and the result was very close to that of Ni-ABCBA-Ni structure. For comparison, we have performed the same calculations for graphene layers in the same stacking but without Ni slabs above. The results of these calculations are presented in Fig.~\ref{fig2}(c).

\begin{figure}
\includegraphics[width=8.5cm]{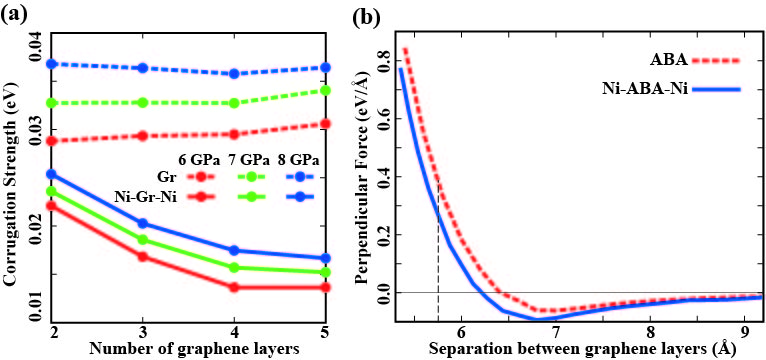}
\caption{(a) Variation of the corrugation strength, $W_D$ with the number of graphene layers, $n$ for three different loading pressures (with and without Ni(111) substrates). (b) Perpendicular force $F_z$ versus the separation distance between outermost graphene layers for Ni-ABA-Ni and ABA structures ($n$=3). In the repulsive range, the perpendicular force and hence the potential corrugation is larger in the absence of Ni(111) slabs.}

\label{fig3}
\end{figure}

Figure~\ref{fig3} presents crucial trends related with the corrugation strength $W_D$. As expected, $W_D$ increases with increasing normal force and is higher in structures composed of only graphene layers (like ABA) compared to the ones having Ni slabs (like Ni-ABA-Ni). This effect is mirrored in the repulsive interaction of graphene layers in the presence and absence of Ni slabs, as shown in Fig.~\ref{fig3} (b). Namely, for a given width of graphene layers, Ni-ABA-Ni is exerted by a normal loading force, which is smaller than that exerts on ABA without supporting Ni slabs. This situation presents a qualitative explanation why the corrugation strength calculated for Ni-ABA-Ni is smaller than that calculated for ABA in Fig.~\ref{fig3} (a). Here one can see that the repulsive force for a given separation of graphene layers is reduced in the presence of Ni slbs. This is consistent with the decrease of $W_D$. Another important finding is that $W_D$ of "AB.." structures solely composed of graphene layers has minor variation with the number of layers. On the other hand, $W_D$ of Ni-AB..-Ni structures decreases gradually with increasing $n$ and eventually saturates at a value for $n >$ 3. This variation of the corrugation strength is in compliance with the  nature of short range chemical interaction between Ni(111) and graphene layers. Stated differently, the exponentially decaying overlap between Ni-$3d$ and graphene-$\pi$ orbitals reflects to the variation of $W_D$ with $n$. Even if this trend illustrated in Fig.~\ref{fig3}(a) is seemingly reminiscent of the experimental observations related with the sliding friction between tip and graphene layers,\cite{lee,carpick} the system at hand is very different and heralds another important effect.

\begin{figure}
\includegraphics[width=8.5cm]{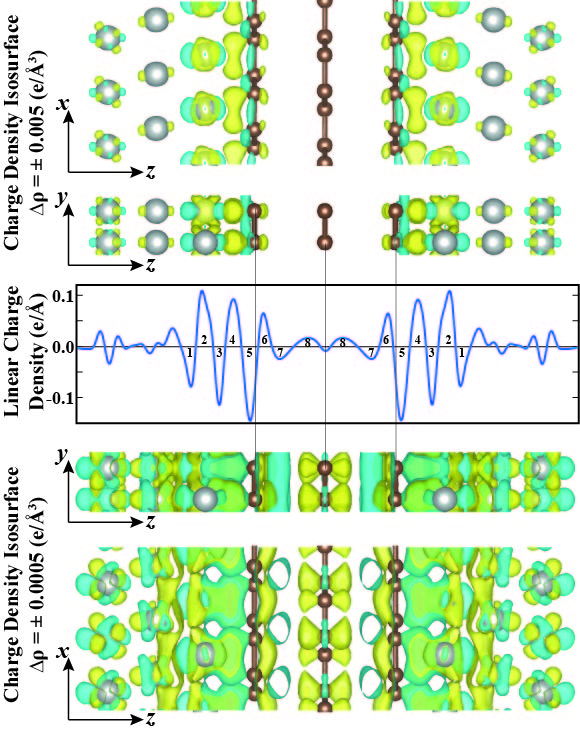}
\caption{Top and bottom panels are isosurfaces of difference charge density, $\Delta \rho$. The bottom panel has isovalue one order of magnitude smaller than that of the top panel. Linear (or planar averaged) $\Delta \rho$ varying along $z$-axis is given in the middle panel with numerals indicating specific regions. The normal pressure is $\sim$~6~GPa. Yellow (blue) isosurface plots correspond to the charge density accumulation (depletion). For the definition of difference charge density $\Delta \rho$ see the tex.}

\label{fig4}
\end{figure}

These trends can be explained by the effects of Ni slabs on the electronic structure of graphene layers. The self-consistent difference charge density $\Delta \rho$, is obtained by subtracting the charge density of ABA structure and two Ni(111) slabs from that of Ni-ABC-Ni structure. The isosurfaces of $\Delta \rho$ and the variation of its value averaged over $(x,y)$-planes parallel to graphene layers (called linear density) are presented in  Fig.~\ref{fig4}. The major charge transfer takes place between Ni and graphene layers attached to each other as seen in the top and middle panels of Fig.~\ref{fig4}. In addition to the analysis of the difference charge density, we performed also the orbital projection analysis of relevant bands of Ni-ABC-Ni. The dangling Ni-$d_{z^2}$ orbitals at the surface of the Ni slab mix with carbon orbitals upon coating by graphene layers. This is resulted in the charge depletion denoted by the numerals 1 and 3 in the linear difference charge density plot. Our analysis of the band structure reveals also significant contribution to C-$p_z$ states from $s$, $d_{xz}$ and $d_{yz}$ orbitals of Ni atoms, while C-$p_z$ orbitals by themselves contribute to $d_{xy}$ and $d_{x^2}$ states of Ni atoms. As a result of these complex mechanism of charge transfer the charge density around the graphene layer is shifted towards Ni slab resulting in charge accumulations (depletions) denoted by numerals 4 and 6 (5 and 7).

The charge density depletion denoted by numeral 7 in $\Delta \rho(z)$  may be the key feature to explain the decrease in the corrugation strength between graphene layers due to Ni slabs. The isosurface of charge depletion corresponding to this region can be seen in the bottom panel of ~\ref{fig4}. \textit{This charge depletion lowers the chemical interaction between graphene layers and results in the lowering of corrugation strength as seen in Fig.~\ref{fig2} and Fig.~\ref{fig3}}. Moreover, similar charge depletions are also observed in Ni-AB-Ni, Ni-ABCA-Ni structures and  their amplitude exponentially decreases by going from two to five layers. This is in accordance with the decrease in the corrugation strength with increasing number of layers, shown in Fig.~\ref{fig3}(b).

\subsection{Discussions}
The recent experimental work,\cite{kim} which has investigated the lubrication capacity of graphene can provide crucial supports for the present theoretical study. It is already known that inert gas atoms or inert materials placed between two flat metal surfaces reduces the adhesion and friction by separating metal surfaces.\cite{persson,buldum} Accordingly, the experimental work by Kim \textit{et. al.}\cite{kim} confirms that graphene multilayers, in fact, decouple the interaction between two insulators, namely SiO$_2$ lens and SiO$_2$ substrate, whereby the adhesion and the coefficient of friction are reduced. The graphene layer and SiO$_2$ surface are neither commensurate, nor pseudomorphic. Therefore, the coating of SiO$_2$ by graphene is vulnerable to rippling and wear. In the present study considering a single layer graphene placed between two flat metal surfaces, the metal-graphene coupling becomes decisive in determining the pull-off force and friction coefficient. In the case of graphene bilayer, top and bottom layers become stuck to nearly lattice matched, bare Ni(111) surfaces. Hence, the binding between single graphene layer and  the Ni(111) surface is significant. This also explains why rather good quality graphene can be grown by CVD on Ni(111) surface.\cite{lahiri,ongun} On the other hand, graphene-graphene coupling is weakened owing to the chemical interaction between Ni-$3d$ and graphene-$\pi$ orbitals revealed in the previous section. For the same reason, the sliding friction of Ni-AA-Ni occurs in the continuous sliding regime, whereby the energy dissipation through the generation of non-equilibrium phonons is suppressed. Therefore, in comparison with graphene deposited SiO$_2$ substrate\cite{kim} the reducing of adhesion and the lowering of friction coefficient between Ni(111) surfaces coated with single layer graphenes are much pronounced and graphene coating of Ni(111) surfaces are more durable. Kim \textit{et. al.}\cite{kim}, who attained the highest durability and lowest friction coefficient in the sliding of SiO$_2$ lens on the as-grown graphene on Ni corroborates our predictions. Additionally, in the present theoretical work the interesting variation of adhesion and friction with $n$ are revealed by presenting all underlying physical and chemical interaction obtained from first-principles calculations.

Next, we address the question whether the static treatment of sliding friction, where dynamic conditions are neglected is appropriate. The criterion for the frictional figures of merit developed earlier theoretically\cite{seymur} and the relevant experimental background\cite{meyer} is quite general and is independent of dynamical conditions. The sudden processes generating non-equilibrium dynamical events ceases if a proper figure of merit is attained. As demonstrated above the figure of merit of Ni-AB..-Ni structures approximately five times larger than the critical value. Nonetheless, even if the stick-slip behavior disappears and the relative motion proceeds in the continuous sliding regime, our analysis considers the dissipation of maximum amount of energy by other mechanisms and hence requires that the integral of all positive work done in Fig.~\ref{fig2}(b) is dissipated. This requirement makes our analysis more stringent and our conclusions more reliable than those can be obtained by considering only the dynamical effects at room temperature. Thus, our conclusions are not affected because the dynamical effects or stochastic defects of real surfaces could not be simulated using first-principles methods.

\section{Conclusions}
In conclusion, even if the strong interaction between the sliding surfaces of Ni(111) is dramatically reduced by a single layer graphene placed in between, the bistability between approach and pull-off remains. Also the stick-slip motion still exists and continues to dissipates significant amount of mechanical energy. The stick-slip motion and hence the generation non-equilibrium phonons are eliminated with the onset of continuous sliding, once each of metal surfaces in relative motion is coated by a single graphene layer. This is attributed to substantial interaction between Ni surface and graphene through complex charge exchange causing to the reduction of the chemical interaction between graphene layers and hence to the decrease of the corrugation strength. The corrugation strength continues to decrease gradually with increasing graphene layer and eventually saturates at a small value. In the absence of metal slabs each coated by a graphene layer, the corrugation strength is relatively higher and practically independent of the number of graphene layers. Our results demonstrate that graphene attached to sliding surfaces operate as superlubricant. One expects to achieve similar lubrication effect but in lesser degree by placing graphene flakes between sliding or rolling Ni(111) surfaces. The interaction between Ni(111) and graphene investigated in this study appears to be important not only for the growth of pristine graphene or for the protection from oxidation, but also for achieving the nearly frictionless friction. Easy growth of graphene on Ni(111) surfaces makes Ni also an attractive substrate for nanotribology applications.

Finally, we note that the first-principles calculations of potential corrugations calculated in the constant force mode are achieved by optimizing atomic structure. This way, the elastic deformations of sliding surfaces under perpendicular loading force are taken into account. We believe that this important feature of the present method will be used in future studies dealing with the development of lubricant single layer materials.

\section{Acknowledgements}
This work is supported by TUBITAK through Grant No:108T234 also by EFS EUROCORE programme FANAS and the Academy of Sciences of Turkey(TUBA). All computational resources have been provided by TUBITAK ULAKBIM, High Performance and Grid Computing Center (TR-Grid e-Infrastructure).

\end{document}